\documentclass[aoas,MSNbibl,nameyear,dvips]{arximspdf}

\doi{10.1214/14-AOAS728C} % kopijuoti is laisko
\volume{8}
\issue{4}
\pubyear{2014}
\firstpage{1956}
\lastpage{1960}
\docsubty{FLA}
\referstodoi{10.1214/14-AOAS728}


\begin{document}
\begin{frontmatter}

\title{Discussion of ``Spatial accessibility of pediatric primary healthcare: Measurement and inference''}
\runtitle{Discussion}

\begin{aug}
\author[A]{\fnms{Lance A.}~\snm{Waller}\corref{}\ead[label=e1]{lwaller@sph.emory.edu}}%,
% \and
\runauthor{L.~A. Waller}
\affiliation{Emory University}
\address[A]{Department of Biostatistics and Bioinformatics\\
Rollins School of Public Health\\
Emory University\\
1518 Clifton Road NE\\
Atlanta, Georgia 30340\\
USA\\
\printead{e1}}
\end{aug}

% HISTORY:
\received{\smonth{6} \syear{2014}}

% ABSTRACT

% KEYWORDS
\end{frontmatter}

\section*{Introduction}

Nobles, Serban and Swann (\citeyear{NobSerSwa14}) provide a thoughtful and thorough
contribution to the literature on modeling and assessing spatial
disparities in access to healthcare. The proposed models include several
features I find particularly attractive, including:
\begin{itemize}
\item A model focusing on structural disparities in access to
healthcare, an important precursor to resulting health disparities. There
is a large literature measuring disparities in health outcomes, and models
such as the authors' allow a framework for assessing potential policy
impacts. The authors' results illustrate the importance of such an exercise
by indicating that some straightforward solutions, for example, simply
increasing the proportion of Medicaid patients accepted within a practice,
may not result in appreciable changes in the access-based disparities.
\item[$\bullet$] The approach includes both an optimization component, to
describe access and healthcare choices, and a statistical component, to
estimate association with socioeconomic measures at the census tract level.
The result is a rich blend of tools from operations research, optimization
and statistics.
\item Finally, the authors avoid the temptation to focus on a
single model, instead summarizing results across statistical models since
the local covariates are often highly collinear.
\end{itemize}

The authors' approach, application and results provide new insights and
point to new directions for future research. I offer thoughts in four areas
(often leaning on the discussant's prerogative of raising rather than
answering questions!): (1)~Assessments of disparities, spatial assessments
of disparities and assessment of spatial disparities, (2) Spatial variation
of covariates, spatial variation of associations, and spatial scale, (3)
Aggregate, local and individual impacts, and (4) Data availability and the
dynamic healthcare environment.

\section*{Assessments of disparities, spatial assessment of disparities and
assessments of spatial disparities}

The authors address an important component of health disparities, namely,
estimation and inference relating to disparity in access to and use of
healthcare. There are extensive, but largely separate, literatures relating
to issues of measuring and analyzing disparities in several different
fields, including the health disparities literature, a robust econometric
literature regarding inference for income disparities [e.g., Gastwirth, Nayak and Wang
(\citeyear{GasNayWan89})], and a literature associated with environmental justice focused
on inference for disparities in environmental exposures [e.g., \citet{Liu01}].
There is little cross-fertilization in methods used between these
application areas, but some similar ideas appear, such as the desire for
methods assessing differences across the full distribution of outcome
within comparison groups (e.g., sex or race) rather than single summary
statistics, for example, using methods such as relative distribution
methods for income distributions [\citet{HanMor99}] or integrated
cumulative distribution functions of exposures in environmental justice
[Waller, Louis and Carlin (\citeyear{WalLouCar})]. It will be interesting to place the authors' work
in this broader context to provide stronger links between the proposed
methodology and variants on the motivating questions in other settings.

With statistical assessments of disparities in place, one next moves to
spatial assessments of local disparities providing maps of local estimates
of disparity (and associated uncertainty). The authors provide an
attractive framework using confidence bands to identify unusual local
variations for further review. Other approaches utilize hierarchical models
based on small area estimation and local smoothing priors to stabilize
local ratio estimates and posterior predictive distributions to assess
probabilities of exceeding given thresholds [Tassone, Waller and Casper (\citeyear{TasWalCas09})]. The
key to both approaches is a statistical assessment of local observed
disparities in order to identify areas with particularly high (or low)
disparity, with a focus on identifying spatial variations in disparity
under current or proposed policies.

In addition to assessments of disparities and spatial assessments of
disparity, there is also a (smaller) literature on a particularly spatial
aspect of disparity, namely, what would the impact on disparity be if
policies changed at a particular location or set of locations. The
literature on equitable facility location [e.g., \citet{MarSch94}, as referenced by the authors]
provides an example, that is, what is
the impact on equitable access to a new facility location placed at a
particular location whether the facility is a societal benefit (e.g., a
library or health clinic) or a detriment (e.g., an environmental hazard)?
Such approaches typically estimate a surface reflecting the resulting
impact on disparity for a facility placed at any particular location across
the study area [Waller, Louis and Carlin (\citeyear{WalLouCar}) and the references therein]. This
surface represents adjustments to disparity based on changes at a
geographic location rather than estimated current level of disparity at a
location, and offers additional insight for evaluating proposed local
changes. In the authors' example, suppose we could add a fixed number of
pediatric primary care clinics to the state, where should we place them in
order to provide the largest improvement in overall spatial accessibility?
There is much room for further growth of a core methodological framework
assessing such changes and it would be interesting to investigate how the
authors' approach can offer insight into this setting.

\section*{Spatial variation of covariates, spatial variation of associations
and spatial scale}

The authors note the difference between spatial variation in covariate
values and the spatial variation of associations between outcomes and
covariates. This distinction is important and merits repeating. In studies
of health disparities, outcome and covariate data are often obtained from
different institutions and instruments. Location provides the link between
covariates and outcomes and, even though values vary by location, the
spatial aspect of modeling can be ignored and associations measured by, for
example, standard (aspatial) regression or generalized linear models. A
particularly spatial challenge is when the associations (model parameters)
vary by location and the authors' approach builds on methods to
statistically map these spatially varying associations, with interesting
results.

In the authors' application, many of these spatially varying effects seem
to hinge on urban--rural differences with variation between the Atlanta
metropolitan area and more rural parts of the state. The authors mention
this distinction, but it may merit further investigation. Conceptually,
different factors will operate at different spatial scales, and these
scales may operate differently for populations in urban tracts than those
in rural tracts. The authors mention distance to care, noting a 25-mile
limit. While this limit may cover most rural areas, some rural areas may
well include primary care clinics more than 25 miles from an individual's
residence, and the rate of distance-decay associated with the gravity model
may be different for individuals in urban than in rural tracts. In
addition, the authors' model synthesis approach may provide room for
additional insight into urban/rural differences. Would it be possible to
assess whether different models are driving results in the urban and rural
tracts? The urban/rural differences may be difficult to model and assess,
but they seem to pervade the results and the structure of the authors'
approach may offer new opportunities for insight into factors driving the
optimization, factors associated with outcomes, and factors likely to
impact policy effects in the urban settings, rural settings or both.

\section*{Aggregate, local and individual impacts}

Accurate local assessments of healthcare disparities provide important
input for defining and evaluating local policies to alleviate these
disparities. The authors' approach provides a structure for estimating
current disparities and the impact of policy changes at the tract level,
especially for policies impacting elements of the optimization component of
the model (e.g., the provider's willingness to accept Medicaid patients).
The approach offers the opportunity to assess and summarize impacts of
changes in these factors at the state, regional, local or individual level.
Regional variations in healthcare policy, even at the federal level, are
challenging but not impossible to implement [see, e.g., a recent \citet{autokey3}
report on geographic variation in Medicare
reimbursement].

It is important to recognize that the authors' analysis (like most analyses
of similar data) is largely observational, linking multiple sources of
geographically referenced data for analysis. As an observational study of
spatially referenced data aggregated to the census tract level, inference
to the individual level can be challenging due to the ``ecologic fallacy''
of epidemiology (individual level associations may differ from associated
observed in aggregate) and the ``modifiable areal unit problem'' of
geography (aggregate associations may differ if individuals are aggregated
into different sets of regions). In addition, the authors note their
targeting of approximately Pareto optimal solutions to improve some
dimensions of accessibility for some groups without significantly reducing
accessibility for others. While well beyond the scope of the current paper,
I wonder about potential links between these epidemiologic, geographic and
optimization issues, all addressing aspects of individual level inference
in aggregate data, and whether we might gain additional understanding by
considering them together.

\section*{Data availability and the dynamic healthcare environment}

The authors' analytic approach depends on data from a variety of sources
and many of these are changing, not only in content but also structure,
accuracy and availability.

The authors provide inference at the census tract level, drawing on
tract-level sociodemographic and economic data from the U.S. Census. In
2010, the American Community Survey (ACS) replaced the Census long form as
the source of tract-level data for many economic variables. The ACS
involves a rolling sample across the U.S., providing many benefits for
national and regional estimates but also challenges in their accurate use
and replacement of long form-based estimates [\citet{Cou07}]. Relevant to the authors' work, Spielman, Folch and Nagle
(\citeyear{SpiFolNag14}) report an
average 75\% increase in uncertainty at the census tract level in ACS
estimates compared to past long form estimates. Spielman, Folch and Nagle (\citeyear{SpiFolNag14}) also
examine observed geographic patterns in this uncertainty and illustrate
measurable associations with local covariates, such as distance to urban
centers. These features suggest a need to incorporate errors-in-covariates
and, perhaps even, spatial modeling of these errors-in-covariates in future
extensions of the authors' work, especially when extending the methods
longitudinally to assess changes in disparity over time to pre- and
post-ACS time periods.

In addition to changes in census data, the healthcare environment is
dynamic, not only at the federal level with the Affordable Care Act, but
also in local and individual reactions to changes in the system. Recent
years have seen changes in healthcare provision (e.g., the rise in ``urgent
care'' facilities), healthcare utilization (e.g., the use of emergency
departments for primary care) and urban/rural differences in these changes.
The authors' focus on pediatric primary care narrows the impact of some of
these changes, but such issues could have impact on extension to broader
elements of healthcare and on longitudinal impacts.

\section*{Summary}

In summary, I thank the authors for a thought-provoking analysis of a very
challenging set of problems. The work provides important insight into its
present application and an analytic framework for continued application in
a challenging and dynamic environment.

% imsref loaded by akundreckaite, 2014-10-03 15:41:51
% imsref loaded by akundreckaite, 2014-10-03 16:16:19


%
%

\printaddresses
\end{document}